\documentclass[12pt]{article}
\usepackage{times}

\def\beq{\begin{equation}}
\def\eeq{\end{equation}}
\def\bea{\begin{eqnarray}}
\def\eea{\end{eqnarray}}

\def\la{\langle}
\def\ra{\rangle}

\def\nn{\nonumber}
\def\Eq#1{Eq.~(\ref{#1})}

\def\x#1{\langle#1\rangle}

\newcommand{\ta}[1]{#1\hspace{-.42em}/\hspace{-.07em}}

\def\tap{\ta{\hat{p}}}

\begin{document} 
%%%%%%%%%%%%%%%%%%%%%%%%%%%%%%%%%%%%%%%%%%%%%%%%%%%%%%%%%%%%%%%%%%%%%

\begin{titlepage}

\renewcommand{\thefootnote}{\fnsymbol{footnote}}
\begin{flushright}
    IFIC/05-37   \\ hep-ph/0508138
\end{flushright}
\par \vspace{10mm}

\begin{center}

{\Large \bf Multigluonic scattering amplitudes of heavy quarks}

\vspace{8mm}

{\bf Germ\'an Rodrigo~$^{(a)}$\footnote{E-mail: german.rodrigo@ific.uv.es}}

\vspace{5mm}

${}^{(a)}$ Instituto de F\'{\i}sica Corpuscular, 
CSIC-Universitat de Val\`encia, \\
Apartado de Correos 22085, 
E-46071 Valencia, Spain. \\

\vspace{5mm}

\end{center}

\par \vspace{2mm}
\begin{center} {\large \bf Abstract} \end{center}
\begin{quote}
\pretolerance 10000
We consider heavy quark and antiquark 
off-shell spinorial currents with emission of an arbitrary number of 
gluons of positive helicity. From this results we calculate the 
corresponding on-shell scattering amplitude and the quark-antiquark 
vector current. Then, we show that 
in the heavy top quark effective theory the holomorfic 
component of the Higgs $\to q \bar{q} + n$-gluon amplitude
vanish for helicity conserving configurations. 
\end{quote}

\vspace*{\fill}
\begin{flushleft}
     IFIC/05-37 \\ August 11, 2005
\end{flushleft}
\end{titlepage}

\setcounter{footnote}{1}
\renewcommand{\thefootnote}{\fnsymbol{footnote}}

%%%%%%%%%%%%%%%%%%%%%%%%%%%%%%%%%%%%%%%%%%%%%%%%%%%%%%%%%%%%%%%%%%%%%

\section{Introduction}

Production of jets at large transverse momentum with respect to 
the beam is the typical signature at high energy hadron colliders, 
both for signal and background. The LHC will produce an enormous 
amount of heavy quarks. Indeed it can be considered a 
heavy quark factory. Since heavy quarks will 
appear associated to jets, the calculation of multipartonic 
scattering amplitudes with heavy fermions is mandatory in order 
to provide accurate phenomenological predictions. 
But the complexity of perturbative calculations increases 
exponentially with the number of jets, in such a way that
the usual expansion in Feynman diagrams becomes prohibitive.  
Recursion relations within the helicity amplitude 
formalism~\cite{Jacob:1959at,Bjorken:1966kh,Mangano:1990by}
have been proven to be an elegant and efficient tool
to calculate multipartonic scattering amplitudes,
and might allow to overcome this problem. Together with the calculation 
of the scattering amplitudes, more efficient methods to integrate 
over the intricate multipartonic phase-space are also required 
for the development of Monte Carlo event generators. 

Recursion relations have been extensively used in the 
literature at tree~\cite{Berends:1987me,Dixon:1996wi}
and one-loop level~\cite{Bern:1994zx,Bern:1994cg} in the past.
Witten's idea of a duality between supersymmetric Yang-Mills 
and topological string theories in twistor space~\cite{Witten:2003nn} 
has reopened the interest in this issue. 
Since then, a new method for the evaluation of scattering amplitudes 
in gauge theories has been proposed~\cite{Cachazo:2004kj}.  
It is based on the recursive use of off-shell 
Maximal Helicity Violating amplitudes (MHV)~\cite{Parke:1986gb}. 
Recent works have accomplished interesting progress since the original 
formulation, and the method has been refined by 
introducing more efficient recursion relations~\cite{Britto:2004ap,Britto:2005fq}, 
and extending this approach to the one-loop level~\cite{Bern:2005cq,Bern:2005hs}. 
Although initially formulated for massless particles the MHV rules 
have been generalized to include heavy particles.
Within this formalism recursion relations for massive scalar particles 
at tree-level have been introduced in Ref.~\cite{Badger:2005zh}
and for vector boson and fermions in Ref.~\cite{Badger:2005jv}.
Those recursion relations have been used by the authors of 
Ref.~\cite{Forde:2005ue} to calculate multigluonic amplitudes 
with heavy scalars. 

Due to the phenomenological impact of heavy quark production at 
LHC, we analyze in this paper the potentiality of the Berends-Giele 
recursion relations~\cite{Berends:1987me} to 
evaluate several helicity amplitudes with heavy quarks and an 
arbitrary number of gluons. We obtain compact expressions for 
some helicity configurations. We find that this method
provides efficiently valuable information for the calculation of 
scattering amplitudes with heavy particles, in a complementary way 
to the MHV approach.

\section{Scattering amplitudes with one quark-antiquark pair}

Spinors for massive fermions can be constructed from two null 
vectors~\cite{Hagiwara:1985yu,Tanaka:1989gu,Ballestrero:1992dv}.
The decomposition is however not unique, and the freedom to choose 
such vectors makes the helicity amplitude method very powerful
also for massive particles. 
We shall consider processes where quark-antiquark pairs are produced.
We denote by $p_1^\mu$ and $p_2^\mu$, with $p_1^2=p_2^2=m^2$, the end 
point four-momentum of the fermion line. In terms of  
two light-like vectors ($\hat{p}_1^2=\hat{p}_2^2=0$)
the outgoing fermion-antifermion four-momenta can be written as 
\bea
p_1^\mu &=& \frac{1+\beta}{2} \, \hat{p}_1^\mu + 
\frac{1-\beta}{2} \, \hat{p}_2^\mu~, \nn \\
p_2^\mu &=& \frac{1-\beta}{2} \, \hat{p}_1^\mu + 
\frac{1+\beta}{2} \, \hat{p}_2^\mu~, 
\eea
where $\beta=\sqrt{1-4m^2/s_{12}}$ is the velocity of the quark, 
and $s_{12} = (p_1+p_2)^2$. This momentum transformation is of 
course not unique. The advantage of our choice of normalization
is that it preserves momentum conservation since trivially 
$\hat{p}_1+\hat{p}_2=p_1+p_2$, and hence 
$\hat{s}_{12}=(\hat{p}_1+\hat{p}_2)^2=s_{12}$.
Furthermore, in the massless limit we have: $p_1\to \hat{p}_1$ 
and $p_2\to \hat{p}_2$. The inverse transformation reads:
\bea
\hat{p}_1^\mu &=&  \frac{1+\beta}{2\beta} \, p_1^\mu - 
\frac{1-\beta}{2\beta} \, p_2^\mu~, \nn \\
\hat{p}_2^\mu &=& - \frac{1-\beta}{2\beta} \, p_1^\mu + 
\frac{1+\beta}{2\beta} \,p_2^\mu~. 
\eea
Then, we use $\hat{p}_1$ and $\hat{p}_2$ as reference 
null vectors to define the quark-antiquark spinors:
\beq
\bar{u}_\pm(p_1,m) = \frac{\beta_+^{-1/2}}{\x{2^\mp | 1^\pm}} 
\langle 2^\mp | \, (\ta{p}_1+m)~, \qquad
v_\pm(p_2,m) = \frac{\beta_+^{-1/2}}{\x{2^\mp | 1^\pm}}
(\ta{p}_2-m) \, |1^\pm \rangle~,
\eeq
where $\beta_\pm=(1\pm\beta)/2$, and we use the shorthand 
notation $| i^\pm \rangle = | \hat{p}_i^\pm \rangle$. With this 
definition the massive spinors have a smooth massless limit.
For $m=0$: 
\beq
\bar{u}_\pm (p_1) = \langle 1^\pm|~, \qquad 
v_\pm (p_2) = | 2^\mp \rangle~.
\eeq

We consider now a quark-antiquark vector current. In the 
helicity basis, we find:
\beq
S^\mu(1^\pm_q,2^\mp_{\bar{q}}) = \bar{u}_\pm(p_1,m) \gamma^\mu v_\mp (p_2,m) 
= \x{1^\pm|\gamma^\mu| 2^\pm}~.
\eeq
Namely, the vector current has the same functional form as 
the massless one but replacing the massless momenta by the 
corresponding null vectors used to define the massive spinors. 
For massive fermions, we shall consider also the case of 
helicity flip along the fermion line, which would vanish 
in the massless limit. We obtain:
\beq
S^\mu(1^\pm_q,2^\pm_{\bar{q}}) = \bar{u}_\pm(p_1,m) \gamma^\mu v_\pm (p_2,m) 
= \frac{2 \, m}{\x{2^\mp| 1^\pm}} \,(\hat{p}_2-\hat{p}_1)^\mu~.
\eeq

Generalizing these results to the insertion of an arbitrary 
odd number of gamma matrices between the spinors, the 
following useful properties hold: 
\bea
S^{\mu_1 \ldots \mu_{2n+1}}(1^\pm_q,2^\mp_{\bar{q}}) &=& 
\bar{u}_\pm(p_1,m) \gamma^{\mu_1} \cdots \gamma^{\mu_{2n+1}} v_\mp (p_2,m) 
\nn \\ &=& 
\beta_+ \x{1^\pm| \gamma^{\mu_1} \cdots \gamma^{\mu_{2n+1}} | 2^\pm}
+ \beta_- \x{1^\pm| \gamma^{\mu_{2n+1}} \cdots \gamma^{\mu_1} | 2^\pm}~,
\label{oddcurrent}
\eea
and 
\bea
S^{\mu_1 \ldots \mu_{2n+1}}(1^\pm_q,2^\pm_{\bar{q}}) &=& 
\bar{u}_\pm(p_1,m) \gamma^{\mu_1} \cdots \gamma^{\mu_{2n+1}} v_\pm (p_2,m) 
\nn \\ &=& \frac{m}{\x{2^\mp| 1^\pm}} \,
\left(\x{2^\mp|\gamma^{\mu_1} \cdots \gamma^{\mu_{2n+1}}|2^\mp}
     -\x{1^\pm|\gamma^{\mu_1} \cdots \gamma^{\mu_{2n+1}}|1^\pm}\right)~.
\label{flipcurrent}
\eea
In the massless limit $\beta_+ \to 1$ and $\beta_- \to 0$, and only 
the first term in \Eq{oddcurrent} survives, while \Eq{flipcurrent}
obviously vanishes.
Similar relations can be derived for the insertion of an even 
number of gamma matrices as well, but those are not needed 
for the current purpose of this paper. 

In the following we will consider the quark-antiquark vector current, 
with emission of an arbitrary number of gluons 
$V^\mu \to q \bar{q}+(n-2) g$. 
The colour decomposition is such that
\beq
\hat{S}^\mu(1_q;2,3,\ldots,n-1;n_{\bar{q}}) = 
g_\mathrm{S}^{n-2} \sum_{P(2,\ldots, n-1)} (T^{a_2}T^{a_3} \cdots T^{a_{n-1}})
\, S^\mu(1_q;2,3,\ldots,n-1;n_{\bar{q}})~,  
\eeq
where $P(2,\ldots,n-1)$ is the permutation group of all the gluons, 
and $T^{a_i}$ are colour matrices in the fundamental representation.

\section{Spinorial off-shell currents}

Recursion relations for the calculation of the spinorial current 
of a quark-antiquark pair and $n$-gluons, where either the quark 
or the antiquark are off-shell, have been derived long time 
ago~\cite{Berends:1987me}. 
Those recursion relations are valid regardless the massive or massless 
character of the quark-antiquark pair, although explicit results where 
obtained only in the massless approximation. In this Section, we 
use these recursion relations to extend their results to the heavy 
quark case. 

The colour ordered spinorial current of an on-shell quark of 
four-momentum $p_1$ and $(m-1)$-gluons of four-momenta $p_2$ to $p_m$ 
is given in terms of the spinorial current of the on-shell quark 
with less gluons, and the off-shell gluonic current $J^\mu$
of the rest of the gluons: 
\beq
S(1_q;2,\ldots,m) = - \sum_{k=1}^{m-1}
S(1_q;2,\ldots,k) \ta{J}(k+1,\ldots,m) \frac{1}{\ta{p}_{1,m}-m}~,
\eeq
where $p_{1,m}=p_1+p_2+\ldots +p_m$ and $S(1_q) = \bar{u}(p_1,m)$. 
For all gluons of positive helicity the gluonic current has the 
form~\cite{Berends:1987me}:
\beq
J^\mu(i^+,\ldots,j^+) = 
\frac{\x{\xi|\gamma^\mu \ta{p}_{i, j}|\xi}}
{\sqrt{2}\x{\xi i}\x{i \,(i+1)}\cdots \x{j\xi}}~.
\eeq
The null vector $\xi$ is the reference gauge vector which is assumed
to be the same for all the gluons. 
Conversely, for the off-shell quark current one has
\beq
S(m+1,\ldots,n-1;n_{\bar{q}}) = \frac{1}{\ta{p}_{m+1, n}+m}
\sum_{k=m+2}^{n} \ta{J}(m+1,\ldots,k-1)
S(k,\ldots,n-1;n_{\bar{q}})~, 
\label{srecursion}
\eeq
the antiquark carrying four-momentum $p_n$, and $S(n_{\bar{q}})=v(p_n,m)$.

It is convenient to work in the gauge where $\xi=\hat{p}_n$, 
where $\hat{p}_n$ is the reference null vector of the antiquark. 
In that gauge, diagrams with gluons of positive helicity attached 
to an antiquark of negative helicity do not contribute in the massless 
approximation. This is not the case for massive quarks, but that 
choice of gauge still provides some simplifications. 
Furthermore: $m^2 = \beta_+ \beta_- \x{1\xi}[\xi1]$.
For the complementary helicity configurations it would be convenient 
to choose $\xi=\hat{p}_1$ instead. 

The no-gluon, one-gluon, and two-gluon spinorial quark currents are
%(factors $\sqrt{2}$)
\beq
S(1_q^+) = \bar{u}_+ (p_1,m) = \frac{\beta_+^{-1/2}}{\x{\xi1}}
\la \xi | (\ta{p}_1+m) = \beta_+^{1/2} \, [1| + 
\beta_+^{-1/2} \, \frac{m}{\x{\xi 1}} \, \la \xi|~,
\eeq
\beq
S(1_q^+;2^+)= - S(1_q^+) \, \ta{\varepsilon}_2^+ \, 
\frac{1}{\ta{p}_{12}-m} = - S(1_q^+) \, 
\frac{|2]}{y_{12} \x{\xi2}} \, \la \xi | (\ta{p}_{12}+m)~, 
\eeq
\bea
S(1_q^+;2^+,3^+)&=& - \left( S(1_q^+;2^+) \ta{\varepsilon}_3^+ 
+ S(1_q^+) \ta{J}(2^+,3^+) \right)
\frac{1}{\ta{p}_{1,3}-m} \nn \\
&=& \frac{S(1_q^+)}{y_{1,3} \, \x{\xi2}\x{\xi3}}
\left( \frac{|2]}{y_{12}} \la \xi|\ta{p}_{12}|3] + 
\frac{\ta{p}_{23}|\xi \ra}{\x{23}} \right) \, 
\la \xi | (\ta{p}_{1,3}+m)~, 
\eea
where we defined $y_{1,n}=p_{1,n}^2-m^2$.
After some algebra we can simplify further the  
two-gluon current
\bea
S(1_q^+;2^+,3^+) &=& - \frac{S(1_q^+)}{\x{23} y_{12}}
\left( \frac{|2]}{\x{\xi3}} + \beta_-
\frac{|\xi] [23] }{y_{1,3}} 
\right) \, \la \xi | (\ta{p}_{1,3}+m)~. 
\eea

Then, we factorize the off-shell current into the form 
\bea
S(1_q^+;2^+,\ldots,n^+) &=& - \frac{S(1_q^+)}{y_{12}}
X(1_q^+;2^+,\ldots,n^+) \, \la \xi | (\ta{p}_{1,n}+m)~,
\label{form}
\eea
where 
\bea
X(1_q^+;2^+) &=& \frac{|2]}{\x{\xi2}}~, \nn \\
X(1_q^+;2^+,3^+) &=& \frac{1}{\x{23}}
\left( \frac{|2]}{\x{\xi3}}
+ \beta_{-} \frac{|\xi] [23]}{y_{1,3}} \right)~.
\eea
The massless limit forces the off-shell massive current to have  
the following structure
\bea
X(1_q^+;2^+,\ldots,n^+) &=& 
\frac{1}{\x{23} \x{34} \ldots \x{(n-1)n}}
\left[ \frac{|2]}{\x{\xi n}} + \beta_{-} \, |\xi] \, A(1_q^+;2^+,\ldots,n^+)
\right]~.
\label{xstructure}
\eea
From \Eq{srecursion}, we can easily derive a recursion relation 
for the $X$ factor of the off-shell current 
\bea
X(1_q^+;2^+,\ldots,n^+) &=&
\frac{1}{\x{\xi n} \, y_{1,n}} \Bigg[
- \frac{\ta{p}_{2,n} |\xi\ra \, y_{12}}
{\x{\xi2}\x{23}\ldots\x{(n-1)n}} \nn \\
&+& \sum_{k=2}^{n-2} X(1_q^+;2^+,\ldots,k^+)
\frac{\x{\xi|\ta{p}_{1,k} \ta{p}_{k+1,n}|\xi}}
{\x{\xi(k+1)} \x{(k+1)(k+2)} \ldots \x{(n-1)n}} \nn \\
&-& X(1_q^+;2^+,\ldots,n-1^+) \la \xi | \ta{p}_{1,n-1}|n] \Bigg]~.
\label{xrecursion}
\eea 
To simplify this recursion relation further and derive an equivalent 
recursion relation for the function $A$, we multiply and divide 
the last term of \Eq{xrecursion} by $\x{(n-1)n}$, such that
\bea
\la \xi | \ta{p}_{1,n-1}|n] \x{(n-1)n} &=& 
\x{n-1 | \ta{p}_n \ta{p}_{1,n-1} |\xi}  = 
\x{n-1 | (\ta{p}_{1,n} - \ta{p}_{1,n-2}) \ta{p}_{1,n})| \xi} \nn \\
&= &\x{n-1 | p_{1,n}^2 - \ta{p}_{1,n-2} 
(\ta{p}_{1,n-2}+\ta{p}_{n-1,n})| \xi} \nn \\
&=& - y_{1,n} \x{\xi (n-1)} - 
\x{n-1 | y_{1,n-2}+\ta{p}_{1,n-2} \ta{p}_{n-1,n} |\xi} ~. 
\label{combine}
\eea
The first term in the rhs of \Eq{combine} generates 
the first term of the rhs of \Eq{xstructure}. Then, 
it is clear that in order to fulfill the massless 
limit, the remaining contributions have to either 
cancel to each other or be proportional to $\beta_-$.
The proof follows by induction. Assuming that 
\Eq{xstructure} is true for $n-1$, we show that 
it is also true for $n$ through the recursion 
relation in \Eq{xrecursion}. For that, we use
recursively the Fierz identity:
\beq
\la k | = \frac{1}{\x{\xi (k-1)}} \left(
\x{k(k-1)} \la \xi | + \x{\xi k} \la k-1|
\right)~,
\eeq
and
\beq
\x{k|y_{1,k}+\ta{p}_{1,k}\ta{p}_{k+1,n}|\xi} =
\x{k|y_{1,k-1}+\ta{p}_{1,k-1}\ta{p}_{k,n}|\xi}~.
\eeq
This results into the following recursion relation for the 
corresponding $A$ function:
\bea
A(1_q^+;2^+,\ldots,n^+) &=&
\frac{1}{\x{\xi n} y_{1,n}} \Bigg[
\sum_{k=2}^{n-1} A(1_q^+;2^+,\ldots,k^+)
\frac{\x{k(k+1)}}{\x{\xi(k+1)}} \x{\xi|\ta{p}_{1,k} \ta{p}_{k+1,n}|\xi} %\nn \\
%&+& A(1_q^+;2^+,\ldots,n-1^+) \x{\xi|\ta{p}_{1,n-1}\ta{p}_n|n-1}
- [2| \ta{p}_{3,n} |\xi \ra \Bigg]~, \nn \\
\label{arecursive}
\eea
where 
\bea
A(1_q^+;2^+)=0, \qquad 
A(1_q^+;2^+,3^+) = \frac{[23]}{y_{1,3}}~. 
\eea
From \Eq{arecursive} we find
\beq
A(1_q^+;2^+,3^+,4^+) = A(1_q^+;2^+,3^+) \left( \frac{\x{\xi3}}{\x{\xi 4}} 
+ \frac{[2|\ta{p}_1 \ta{p}_{23}|4]}
{[23] \,y_{1,4}} \right)~.
\eeq
To obtain this result we have transformed 
he last term in the rhs of \Eq{arecursive} 
in the following way
\beq
[2| \ta{p}_{3,n} |\xi \ra 
= A(1_q^+;2^+,3^+) \frac{[2| y_{1,3}\ta{p}_{3,n} |\xi \ra}{[23]} = 
A(1_q^+;2^+,3^+) \left(\frac{[2| \ta{p}_{1}\ta{p}_{23} \ta{p}_{4,n} | 
\xi \ra}{[23]} + \x{3|\ta{p}_{12}\ta{p}_{3,n}|\xi}
\right)~.
\eeq

Assuming that for an arbitrary number of gluons larger than two
the $A$ function has the form 
\beq
A(1_q^+;2^+, \ldots,n^+) = A(1_q^+;2^+,3^+) \left( \frac{\x{\xi3}}{\x{\xi n}} 
+ \frac{\bar{A}(1_q^+;2^+, \ldots,n^+)}{[23]} \right)~,
\eeq
the $\bar{A}$ function fulfills the recursion relation
\bea
\bar{A}(1_q^+;2^+,\ldots,n^+) &=&
\frac{1}{\x{\xi n} y_{1,n}} \Bigg[
\sum_{k=4}^{n-1} \bar{A}(1_q^+;2^+,\ldots,k^+)
\frac{\x{k(k+1)}}{\x{\xi(k+1)}} \x{\xi|\ta{p}_{1,k} \ta{p}_{k+1,n}|\xi} 
- [2| \ta{p}_{1}\ta{p}_{23} \ta{p}_{4,n} | \xi \ra
\Bigg]~. \nn \\
\label{abarrecursive}
\eea
By solving this equation recursively, we get finally
\bea
&&\!\!\!\!\!\!\!\!\!\!\!\!\!\!\!\!\!\!\!\!\!\!\!\!\!\!\!\!
\bar{A}(1_q^+;2^+,\ldots,n^+) = \sum_{j=1}^{n-4} 
\frac{\bar{A}(1_q^+;2^+,3^+,4^+) \x{45} 
\x{\xi | \ta{p}_{1,4}\ta{p}_{5,w_1} | \xi } -
\x{\xi5} [2|\ta{p}_1 \ta{p}_{23} \ta{p}_{4,w_1} | \xi \ra}
{y_{1,w_1}\cdots y_{1,w_j} } \nn \\
&\times&
\frac{\x{w_1(w_1+1)}\cdots \x{w_{j-1}(w_{j-1}+1)}
\x{\xi | \ta{p}_{1,w_1}\ta{p}_{w_1+1,w_2} | \xi } \cdots
\x{\xi | \ta{p}_{1,w_{j-1}}\ta{p}_{w_{j-1}+1,w_j} | \xi }}
{\x{\xi5} \x{\xi w_1}\x{\xi (w_1+1)} \cdots
\x{\xi w_{j-1}}\x{\xi (w_{j-1}+1)} \x{\xi n}}~, \nn \\
\label{thesolution}
\eea
where $w_k \, \epsilon \, \{5,n\}$, and $w_1<w_2<\ldots < w_j$ with 
$w_j=n$. This completes the calculation of the off-shell quark 
spinorial current. 

\subsection{Antiquark off-shell current}

Very similar results can be obtained for the 
antiquark spinorial current when the antiquark carries
negative helicity. The no-gluon, one-gluon and 
two-gluon currents are given by
\beq
S(n_{\bar{q}}^-) = \beta_+^{1/2} \, |\xi \ra + 
\beta_+^{-1/2} \, \frac{m}{[1\xi]} \, |1]
= - \frac{\beta_+^{1/2}}{m} (\ta{p}_n-m) | \xi \ra~, 
\eeq
\beq
S(2^+;3_{\bar{q}}^-) =  (\ta{p}_{23}-m) |\xi \ra \,
\frac{[2|}{\x{\xi 2}y_{23}}  S(3_{\bar{q}}^-)~,
\eeq
\beq
S(2^+,3^+;4_{\bar{q}}^-) =  (\ta{p}_{2,4}-m) |\xi \ra \,
\left( \frac{[3|}{\x{\xi2}} - \beta_+
\frac{[23][\xi| }{y_{2,4}} 
\right) \, \frac{S(4_{\bar{q}}^-)}{\x{23} y_{34}}~.
\eeq
For an arbitrary number of gluons, we have in general 
\bea
S(k^+,\ldots,n-1^+;n_{\bar{q}}^-) &=& 
(\ta{p}_{k,n}-m) |\xi \ra \,
X(k^+,\ldots,n-1^+;n_{\bar{q}}^-) \, 
\frac{S(n_{\bar{q}}^-)}{y_{n-1,n}}~,
\label{antiform}
\eea
with
\bea
X(k^+,\ldots,n-1^+;n_{\bar{q}}^-) &=& 
\frac{1}{\x{k(k+1)}  \cdots \x{(n-2)(n-1)}} \nn 
\\ &\times&
\left[ \frac{[n-1|}{\x{\xi k}} - \beta_{+} \, [\xi| 
\, A(k^+,\ldots,n-1^+;n_{\bar{q}}^-)
\right]~,
\label{antixstructure}
\eea
where
\beq
A(k^+,\ldots,n-1^+;n_{\bar{q}}^-) =
A(n-2^+,n-1^+;n_{\bar{q}}^-) 
\left( \frac{\x{\xi(n-2)}}{\x{\xi k}} 
+ \bar{A}(k^+,\ldots,n-1^+;n_{\bar{q}}^-) \right)~.
\eeq
A recursion relation similar to \Eq{abarrecursive} can also 
be obtained for this antiquark current, leading to an 
expression similar to \Eq{thesolution} where now the 
labeling of the four-momenta runs backwards.

\section{On-shell scattering amplitudes}

From the spinorial off-shell current one can calculate the amplitude for
the process of $n-2$ outgoing gluons and a quark-antiquark pair. 
This amplitude is obtained from the quark-gluon spinorial current
with $n-2$ gluons, by removing the propagator of the off-shell
quark, contracting the current with the antiquark spinor and 
imposing momentum conservation
\beq
S(1_q^+;2^+,\ldots,n-1^+;n_{\bar{q}}^\mp) = 
- \frac{y_{1,n-1}}{y_{12}} S(1_q^+) X(1_q^+;2^+,\ldots,n-1^+) \la \xi | 
S(n_{\bar{q}}^\mp)\bigg|_{p_{1,n}=0}~.
\label{onshell}
\eeq
The helicity conserving amplitude annihilates, and for the helicity 
flip amplitude we get
\beq
S(1_q^+;2^+,\ldots,n-1^+;n_{\bar{q}}^+) = \frac{m^3}{\beta_+ y_{12} } 
\, \frac{y_{1,n-1} A(1_q^+;2^+,\ldots,n-1^+)}
{\x{23} \x{34} \cdots \x{(n-2)(n-1)} \x{n1}} \Bigg|_{y_{1,n-1}=0}~.
\eeq
In particular
\beq
S(1_q^+;2^+,3^+;4_{\bar{q}}^+) = 
\frac{m^3}{\beta_+ y_{12}} \, \frac{[23]}{\x{23} \x{41}}~.
\eeq

\section{Quark-antiquark vector current}

\label{vectorsection}

From the results obtained in the previous sections we can 
also calculate the off-shell quark-antiquark vector current
\beq
S^\mu(1_q;2,\ldots,n-1;n_{\bar{q}}) = 
\sum_{k=1}^{n-1} S(1_q;2,\ldots,k) \gamma^\mu S(k+1,\ldots,n-1;n_{\bar{q}})~.
\label{vector}
\eeq
For all gluons of positive helicity all the terms contributing 
to the sum in \Eq{vector} have the form 
\beq
S^\mu_k(1_q^+;2^+,\ldots,n-1^+;n_{\bar{q}}^\mp) \propto
\x{\xi| (\ta{p}_{1,k}+m) \gamma^\mu (\ta{p}_{k+1,n}-m)|\xi^\pm}~. 
\eeq
Then, after some algebra we find that
\bea
S^\mu_k(1_q^+;2^+,\ldots,n-1^+;n_{\bar{q}}^-) &\propto&
- m \x{\xi| \tap_{1,n-1} \gamma^\mu |\xi}~, \label{sums} \\
S^\mu_k(1_q^+;2^+,\ldots,n-1^+;n_{\bar{q}}^+) &\propto&
\la \xi| \ta{p}_{1,k} \gamma^\mu \ta{p}_{k+1,n}|\xi] - 2 m^2 \xi^\mu~, 
\eea
Particularly interesting is the helicity conserving current. 
In that case all the terms are proportional to the same quantity,
where now the quark four-momentum has been replaced 
by its reference null vector $\hat{p}_1$. 
Since in \Eq{sums} only massless vectors appear some properties 
that hold for massless scattering amplitudes can be 
generalized to the massive case directly 
without further proof. For some applications 
see Section~\ref{higgs}. We obtain the following expression 
for the helicity conserving vector current:
\bea
&& \!\!\!\!\!\!\!\!\!\!\!\!\!\!\!\!\!\!
S^\mu(1_q^+;2^+,\ldots,n-1^+;n_{\bar{q}}^-) =
\frac{1}{\x{23}\x{34}\cdots\x{(n-2) (n-1)}} %\nn \\ &\times& 
\Bigg\{ 
\frac{\beta_+ [12]}{\x{\xi (n-1)}y_{12}} 
- \frac{\beta_- [1 (n-1)]}{\x{\xi 2} y_{n-1,n}} \nn \\ && 
+ \frac{m^2}{\x{\xi 1}} \left[ \frac{1}{y_{12}} A(1_q^+;2^+,\ldots,n-1^+) 
+\frac{1}{y_{n-1,n}} A(2^+,\ldots,n-1^+;n_{\bar{q}}^-)\right] \nn \\ &&
- \frac{1}{y_{12} \, y_{n-1,n}} \, \frac{m^2}{[1\xi]}
\sum_{k=2}^{n-2} \x{k(k+1)}
\left[ \frac{[12]}{\x{\xi k}} 
+ \beta_- [1\xi] A(1_q^+;2^+,\ldots,k^+) \right] \nn \\ && \times
 \left[ \frac{[(n-1)1]}{\x{\xi (k+1)}} 
+ \beta_+ [1\xi] A(k+1^+,\ldots,n-1^+;n_{\bar{q}}^-)\right] 
\Bigg\} \x{\xi| \tap_{1,n-1} \gamma^\mu |\xi}~.
\label{ngluoncons}
\eea
In the massless limit, the current in \Eq{ngluoncons} agrees 
with the Berends-Giele current~\cite{Berends:1987me}:
\beq
S^\mu(1_q^+;2^+,\ldots,n-1^+;n_{\bar{q}}^-)\bigg|_{m=0} = 
\frac{\x{\xi| \ta{p}_{1,n-1} \gamma^\mu |\xi}}
{\x{12}\x{23}\cdots \x{(n-1) \xi}}~.
\eeq
In the following we present some simple examples.

\subsection{Single gluon emission}

Let us consider a quark-antiquark vector current with emission 
of one single gluon off the quark-antiquark pair.  
For the helicity conserving vector current, we find:  
\bea
S^\mu(1^+_q;2^+;3^-_{\bar{q}}) &=& 
\left( \frac{\beta_+}{y_{12}} - \frac{\beta_-}{y_{23}}\right)
\frac{[12]}{\x{\xi 2}} \x{\xi|\ta{\hat{p}}_{12} \gamma^\mu|\xi} 
= \beta \, \frac{[1|2|\xi \ra }{y_{12} \, y_{23}}
\x{\xi|\ta{\hat{p}}_{12} \gamma^\mu|\xi}~.
\label{onegluoncons}
\eea
The helicity flip configuration is also quite simple, we get 
\beq
S^\mu(1^+_q;2^+;3^+_{\bar{q}}) = m 
\Bigg[ \left( \frac{\beta_+}{y_{12}} - \frac{\beta_-}{y_{23}}\right)
\frac{[12]}{\x{\xi 2}}
\,2 (\hat{p}_3-\hat{p}_1)^\mu + \frac{1}{\x{\xi1}}
[2 | \left( \frac{\tap_1}{y_{12}}
+ \frac{\tap_3}{y_{23}} \right) \gamma^\mu | 2 \ra \Bigg]~,
\label{onegluonflip}
\eeq
which clearly annihilates in the massless limit. 

\subsection{Two gluons vector current}

The helicity conserving vector current with emission of 
two gluons is 
\bea
S^\mu(1^+_q;2^+,3^+;4^-_{\bar{q}}) &=& 
\Bigg[ \frac{1}{\x{23}}\left( 
  \frac{[12] \,\beta_+}{\x{\xi3} \, y_{12}} 
- \frac{\beta_- [13]}{y_{34} \, \x{\xi2}} \right) 
- \frac{m^2}{y_{12} \, y_{34}} \frac{[21][31]}{\x{\xi2}\x{\xi3} [1 \xi]} \nn \\
&-& \frac{m^2 [32]}{\x{23}\x{\xi1}} 
\left( \frac{1}{y_{12} \, y_{1,3}} + \frac{1}{y_{34} \, y_{2,4}}\right)
\Bigg]
\x{\xi|\ta{\hat{p}}_{123} \gamma^\mu|\xi}~.
\eea

\section{Higgs production together with heavy quarks}

\label{higgs}

One of the main objectives of the LHC program is the discovery of
the Higgs boson. Gluons couple to Higgs through a top quark loop 
dominantly, where the top can be integrating out leading to an 
effective interaction. MHV rules for Higgs plus multiparton amplitudes 
have been presented in Refs.~\cite{Badger:2004ty,Dixon:2004za}
in that effective theory. The naive off-shell continuation of 
the Higgs vertices depends of the gauge spinor however, and therefore 
do not work as MHV amplitudes. This problem is 
solved by splitting the Lagrangean into a self-dual and 
an anti-self-dual part, and introducing a complex 
scalar where the Higgs is one of its components
$\phi=\frac12 (H+i A)$. 

In the heavy top 
quark effective theory amplitudes of $\phi$-scalars 
coupled to a quark-antiquark pair and any number of gluons of 
positive helicity vanish~\cite{Dixon:2004za} 
\beq
A_n(\phi,q_1^\pm,g_2^+,\ldots,g_{n-1}^+,\bar{q}_n^\mp)=0~.
\eeq
In this section we shall extend this result to the heavy quark
case. The proof follows almost straitforward from 
the amplitudes that we have calculated in 
Section~\ref{vectorsection}. The helicity conserving 
vector current of a heavy quark-antiquark pair and 
$m-2$ gluons is 
\beq
S^\mu(1_q^+,2^+,\ldots,m-1^+,n_{\bar{q}}^-) = N_{2,m-1}
\x{\xi| \tap_{1,m-1} \gamma^\mu|\xi}~,
\label{vectorhiggs}
\eeq
and the vector current of all gluons of positive 
helicity~\cite{Berends:1987me} reads
\beq
J^\mu(m^+,\ldots,n-1^+) = N_{m,n-1}
\x{\xi|\gamma^\mu \ta{p}_{m,n-1}|\xi}~.
\label{gluonhiggs}
\eeq
The normalization factors $N$ are irrelevant for the present 
proof. What is important is that \Eq{vectorhiggs} has the same
functional form in the massive and in the massless case, where
$\hat{p}_1$ is the null vector used to define the quark spinor. 
The 3-point $\phi g g$ vertex of two off-shell gluons of momenta 
$k_1$ and $k_2$ and Lorentz indices $\mu_1$ and $\mu_2$
is given by \cite{Dixon:2004za}:
\beq
V^\phi_{\mu_1 \mu_2} (k_1,k_2) = g_{\mu_1 \mu_2} \, k_1\cdot k_2 
- k_{1 \mu_2} k_{2 \mu_1} + i \varepsilon_{\mu_1\mu_2\nu_1\nu_2}
k_1^{\nu_1} k_2^{\nu_2}~.
\eeq
Then, one can compute $A_n(\phi,q_1^+,g_2^+,\ldots,g_{n-1}^+,\bar{q}_n^-)$ 
by contracting the vertex with the currents in \Eq{vectorhiggs}
and \Eq{gluonhiggs}, where $k_1=p_{1,m-1}+p_n = \hat{p}_{1,m-1}+\hat{p}_n$,
and $k_2=p_{m,n-1}$.

The 3-point $\phi gg$ vertex contracted with the quark-antiquark and 
the Berends-Giele currents gives 
\bea
V^\phi_{\mu \nu} S^{\mu} J^{\nu}
&=& (k_1 \cdot k_2) (S\cdot J) - (k_1\cdot J)(k_2\cdot S) \nn \\
&+& \frac{1}{4} \sum_{i=m}^{n-1} 
\left( \la i | \gamma^{\mu} \gamma^{\nu} \ta{k}_1|i] +
[i| \gamma^{\mu} \gamma^{\nu} \ta{k}_1|i \ra 
\right) S_{\mu} J_{\nu}~.
\eea
Fierzing the gluonic current
\beq
\ta{J} = 2 N_{m,n-1} \sum_{j=m}^{n-1} \left( |j]\la\xi| 
+ |\xi\ra[j| \right) \x{j\xi},
\eeq
and after some algebra, we find
\beq
V^\phi_{\mu \nu} S^{\mu} J^{\nu} = N_{m,n-1} \, \frac{k_2^2}{2} \, 
\x{\xi| [\ta{S}, \ta{k}_1] |\xi}~,
\label{condition3gg}
\eeq
where $[\ta{S}, \ta{k}_1]$ is the commutator of the quark-antiquark  
current with its total momentum. Fierzing the quark-antiquark 
current
\beq
\ta{S} = 2 N_{2,m-1}\sum_{i=1}^{m-1} \left( |i]\la\xi| 
+ |\xi\ra[i| \right) \x{\xi i},
\eeq
the 3-point vertex obviously cancels since in \Eq{condition3gg}: 
$\la \xi| \ta{S} = 0$ and $\ta{S}|\xi\ra = 0$.

\section{Conclusions}

We have calculated several tree-level multigluonic helicity amplitudes 
with heavy quarks. Our approach is based on the Berends-Giele recursive 
relations. We have shown that the method can be easily applied for the 
evaluation of multipartonic scattering amplitudes with heavy particles,
and leads to compact results. We have considered scattering amplitudes 
with all gluons of the same helicity only, but other helicity 
configurations can be calculated efficiently as well.
Furthermore, interesting phenomenological implications can be derived 
from the general structure of multipartonic amplitudes.

\section*{Acknowledgements}

This work has been supported by Ministerio de Educaci\'on y Ciencia (MEC) 
under grant FPA2004-00996 (PARSIFAL),
Acciones Integradas DAAD-MEC (contract HA03-164),
and Generalitat Valenciana (GV05-015, GV04B-594 and GRUPOS03/013).

\appendix

\section{Spinors and polarization vectors}

Massive spinors of four-momentum $p^\mu$ and mass $m$ can be
constructed with the help of two null vectors $\hat{p}^\mu$ and 
$\hat{q}^\mu$ as follows:
\bea
u_\pm(p,m) &=& \frac{N}{\x{\hat{p}^\mp | \hat{q}^\pm}}
(\ta{p}+m) \, |\hat{q}^\pm \rangle~, \qquad
\bar{u}_\pm(p,m) = \frac{N}{\x{\hat{q}^\mp | \hat{p}^\pm}} 
\langle \hat{q}^\mp | \, (\ta{p}+m)~, \nn \\
v_\pm(p,m) &=& \frac{N}{\x{\hat{p}^\mp | \hat{q}^\pm}}
(\ta{p}-m) \, |\hat{q}^\pm \rangle~, \qquad
\bar{v}_\pm(p,m) = \frac{N}{\x{\hat{q}^\mp | \hat{p}^\pm}} 
\langle q^\mp | \, (\ta{p}-m)~,
\eea
where $N=(p\cdot \hat{q}/\hat{p}\cdot \hat{q})^{-1/2}$
is the normalization factor.
Those spinors trivially fulfill the Dirac equation: 
\beq
(\ta{p} - m) \, u(p,m) = 0~,  \qquad (\ta{p} + m) \, v(p,m)= 0~,
\eeq
and the completeness relation
\beq
\sum_{\lambda=\pm} u_{-\lambda} \bar{u}_\lambda = \ta{p} + m~, \qquad
\sum_{\lambda=\pm} v_{-\lambda} \bar{v}_\lambda = \ta{p} - m~. 
\eeq

The polarization vector of an outgoing gluon of momentum 
$k^\mu$ is defined as
\beq
\varepsilon^\pm_\mu(k,\xi) = \pm 
\frac{\x{\xi^\mp|\gamma_\mu|k^\mp}}{\sqrt{2}\x{\xi^\mp|k^\pm}}~,
\eeq
where $\xi^\mu$ is an arbitrary gauge reference momentum 
satisfying $\xi^2=0$. The polarization vector contracted 
with the gamma matrices reads as follows: 
\beq
\ta{\varepsilon}^\pm(k,\xi) = \pm 
\frac{\sqrt{2}}{\x{\xi^\mp|k^\pm}}
\left( |k^\mp \rangle \langle \xi^\mp| +
|\xi^\pm \rangle \langle k^\pm |\right)~.
\eeq
Other practical properties helping to simplify calculations are 
\beq
\langle i^\mp| \,\ta{\varepsilon}^\pm(k,\xi) \,\ta{k} = 0~, \qquad  
\ta{k} \,\ta{\varepsilon}^\pm(k,\xi) \, | i^\pm \rangle = 0~,
\eeq
and 
\beq
\langle \xi^\mp | \ta{\varepsilon}_\pm(k,\xi)  = 0~, \qquad
\ta{\varepsilon}_\pm(k,\xi) |\xi^\pm\rangle = 0~.
\eeq
Finally, if we choose the reference momenta for all gluons to 
be the same, then it follows that
\beq
\varepsilon^\pm (k_i,\xi) \cdot \varepsilon^\pm (k_j.\xi)=0~.
\eeq


\begin{thebibliography}{90}

%%%% introduction of helicity methods 

\bibitem{Jacob:1959at}
  M.~Jacob and G.~C.~Wick,
  %``On The General Theory Of Collisions For Particles With Spin,''
  Annals Phys.\  {\bf 7} (1959) 404
  [Annals Phys.\  {\bf 281} (2000) 774].
  %%CITATION = APNYA,7,404;%%

\bibitem{Bjorken:1966kh}
  J.~D.~Bjorken and M.~C.~Chen,
  %``High-Energy Trident Production With Definite Helicities,''
  Phys.\ Rev.\  {\bf 154} (1966) 1335.
  %%CITATION = PHRVA,154,1335;%%

%%% report 

\bibitem{Mangano:1990by}
  M.~L.~Mangano and S.~J.~Parke,
  %``Multiparton Amplitudes In Gauge Theories,''
  Phys.\ Rept.\  {\bf 200} (1991) 301.
  %%CITATION = PRPLC,200,301;%%

%%%% recursion relations

\bibitem{Berends:1987me}
  F.~A.~Berends and W.~T.~Giele,
  %``Recursive Calculations For Processes With N Gluons,''
  Nucl.\ Phys.\ B {\bf 306} (1988) 759.
  %%CITATION = NUPHA,B306,759;%%

\bibitem{Dixon:1996wi}
  L.~J.~Dixon,
  %``Calculating scattering amplitudes efficiently,''
  arXiv:hep-ph/9601359.
  %%CITATION = HEP-PH 9601359;%%

\bibitem{Bern:1994zx}
  Z.~Bern, L.~J.~Dixon, D.~C.~Dunbar and D.~A.~Kosower,
  %``One loop n point gauge theory amplitudes, unitarity and collinear limits,''
  Nucl.\ Phys.\ B {\bf 425}, 217 (1994)
  [arXiv:hep-ph/9403226].
  %%CITATION = HEP-PH 9403226;%%

\bibitem{Bern:1994cg}
  Z.~Bern, L.~J.~Dixon, D.~C.~Dunbar and D.~A.~Kosower,
  %``Fusing gauge theory tree amplitudes into loop amplitudes,''
  Nucl.\ Phys.\ B {\bf 435}, 59 (1995)
  [arXiv:hep-ph/9409265].
  %%CITATION = HEP-PH 9409265;%%

%%%%

\bibitem{Witten:2003nn}
  E.~Witten,
  %``Perturbative gauge theory as a string theory in twistor space,''
  Commun.\ Math.\ Phys.\  {\bf 252} (2004) 189
  [arXiv:hep-th/0312171].
  %%CITATION = HEP-TH 0312171;%%

\bibitem{Cachazo:2004kj}
  F.~Cachazo, P.~Svrcek and E.~Witten,
  %``MHV vertices and tree amplitudes in gauge theory,''
  JHEP {\bf 0409} (2004) 006
  [arXiv:hep-th/0403047].
  %%CITATION = HEP-TH 0403047;%%

%%%% MHV

\bibitem{Parke:1986gb}
  S.~J.~Parke and T.~R.~Taylor,
  %``An Amplitude For N Gluon Scattering,''
  Phys.\ Rev.\ Lett.\  {\bf 56} (1986) 2459.
  %%CITATION = PRLTA,56,2459;%%

%%%%

\bibitem{Britto:2004ap}
  R.~Britto, F.~Cachazo and B.~Feng,
  %``New recursion relations for tree amplitudes of gluons,''
  Nucl.\ Phys.\ B {\bf 715} (2005) 499
  [arXiv:hep-th/0412308].
  %%CITATION = HEP-TH 0412308;%%

\bibitem{Britto:2005fq}
  R.~Britto, F.~Cachazo, B.~Feng and E.~Witten,
  %``Direct proof of tree-level recursion relation in Yang-Mills theory,''
  Phys.\ Rev.\ Lett.\  {\bf 94} (2005) 181602
  [arXiv:hep-th/0501052].
  %%CITATION = HEP-TH 0501052;%%

\bibitem{Bern:2005cq}
  Z.~Bern, L.~J.~Dixon and D.~A.~Kosower,
  %``Bootstrapping multi-parton loop amplitudes in QCD,''
  arXiv:hep-ph/0507005.
  %%CITATION = HEP-PH 0507005;%%

\bibitem{Bern:2005hs}
  Z.~Bern, L.~J.~Dixon and D.~A.~Kosower,
  %``On-shell recurrence relations for one-loop QCD amplitudes,''
  Phys.\ Rev.\ D {\bf 71} (2005) 105013
  [arXiv:hep-th/0501240].
  %%CITATION = HEP-TH 0501240;%%

%%%% massive particles 

\bibitem{Badger:2005jv}
  S.~D.~Badger, E.~W.~N.~Glover and V.~V.~Khoze,
  %``Recursion relations for gauge theory amplitudes with massive vector bosons
  %and fermions,''
  arXiv:hep-th/0507161.
  %%CITATION = HEP-TH 0507161;%%

\bibitem{Badger:2005zh}
  S.~D.~Badger, E.~W.~N.~Glover, V.~V.~Khoze and P.~Svrcek,
  %``Recursion relations for gauge theory amplitudes with massive particles,''
  arXiv:hep-th/0504159.
  %%CITATION = HEP-TH 0504159;%%

\bibitem{Forde:2005ue}
  D.~Forde and D.~A.~Kosower,
  %``All-multiplicity amplitudes with massive scalars,''
  arXiv:hep-th/0507292.
  %%CITATION = HEP-TH 0507292;%%

%%%% massive fermions

\bibitem{Hagiwara:1985yu}
  K.~Hagiwara and D.~Zeppenfeld,
  %``Helicity Amplitudes For Heavy Lepton Production In E+ E- Annihilation,''
  Nucl.\ Phys.\ B {\bf 274} (1986) 1.
  %%CITATION = NUPHA,B274,1;%%

\bibitem{Tanaka:1989gu}
  H.~Tanaka,
  %``A Method For Numerical Calculations Of Helicity Amplitudes For Processes
  %Involving Massive Fermions,''
  Comput.\ Phys.\ Commun.\  {\bf 58} (1990) 153.
  %%CITATION = CPHCB,58,153;%%

\bibitem{Ballestrero:1992dv}
  A.~Ballestrero, E.~Maina and S.~Moretti,
  %``Heavy quarks and leptons at e+ e- colliders,''
  Nucl.\ Phys.\ B {\bf 415} (1994) 265
  [arXiv:hep-ph/9212246].
  %%CITATION = HEP-PH 9212246;%%
%\bibitem{Ballestrero:1994jn}
  A.~Ballestrero and E.~Maina,
  %``A New method for helicity calculations,''
  Phys.\ Lett.\ B {\bf 350} (1995) 225
  [arXiv:hep-ph/9403244].
  %%CITATION = HEP-PH 9403244;%%

%%%% Higgs 

\bibitem{Badger:2004ty}
  S.~D.~Badger, E.~W.~N.~Glover and V.~V.~Khoze,
  %``MHV rules for Higgs plus multi-parton amplitudes,''
  JHEP {\bf 0503} (2005) 023
  [arXiv:hep-th/0412275].
  %%CITATION = HEP-TH 0412275;%%

\bibitem{Dixon:2004za}
  L.~J.~Dixon, E.~W.~N.~Glover and V.~V.~Khoze,
  %``MHV rules for Higgs plus multi-gluon amplitudes,''
  JHEP {\bf 0412} (2004) 015
  [arXiv:hep-th/0411092].
  %%CITATION = HEP-TH 0411092;%%



\end{thebibliography}
\end{document}